\documentclass[12pt]{iopart}
\newcommand{\JPSI}{{$J/\psi$}}
\newcommand{\SNN}{{$\sqrt{s_{NN}}$}}
\newcommand{\EE}{{$e^{+}e^{-}$}}
\newcommand{\MUMU}{{$\mu^{+}\mu^{-}$}}
\newcommand{\RAA}{{$R_{AA}$}}

\usepackage[dvips]{graphicx}
\begin{document}
 
\title[\scriptsize{Centrality Dependence of \JPSI~Production in Au+Au and Cu+Cu Collisions by the PHENIX Experiment at RHIC}]{Centrality Dependence of \JPSI~Production in Au+Au and Cu+Cu Collisions by the PHENIX Experiment at RHIC} 
\author{T.~Gunji (for the PHENIX\footnote[7]{{For the full list of PHENIX authors and acknowledgements, see Appendix 'Collaborations' of this volume}} Collaboration)}
 
\address{Center for Nuclear Study, Graduate School of Science, 
  the University of Tokyo, 7-3-1 Hongo, Bunkyo-ku, Tokyo, 113-0033, JAPAN}
\ead{gunji@phenix.cns.s.u-tokyo.ac.jp}
\begin{abstract}
  \JPSI~production has been measured in Au+Au and Cu+Cu 
  collisions at \SNN~= 200 GeV 
  by the PHENIX experiment at the Relativistic Heavy Ion Collider (RHIC) 
  during 2004 and 2005, respectively, at mid-rapidity 
  ($|\eta| \le$ 0.35) via \JPSI~$\rightarrow$ \EE~decay and 
  at forward rapidity (1.2 $\le | \eta | \le$ 2.2) 
  via \JPSI~$\rightarrow$ \MUMU~decay.
  The nuclear modification factor ($R_{AA}$) of $J/\psi$ is presented
  as a function of the collision centrality for Au+Au collisions (final
  results) and Cu+Cu collisions (preliminary results) in both rapidity
  windows. These results are compared to SPS results at lower energy
  and to various theoretical calculations.
\end{abstract}

\section{Introduction}
Heavy quarkonia ($J/\psi$, $\psi^{\prime}$, $\chi_{c}$ and $\Upsilon$) 
has long been considered as one of the most promising probes for 
the deconfinement of the hot and dense QCD medium.
In the deconfined medium,  above a critical temperature $T_c$, 
the yield of heavy quarkonia is predicted to be 
suppressed due to the dynamical color screening effect~\cite{bib:1}.
The dissociation temperature depends on the binding energy of quarkonia
and is extracted to be $\sim$2$T_{c}$ for $J/\psi$ and 
$\sim$1.1$T_{c}$ for $\psi^{\prime}$ and $\chi_c$ from 
quenched lattice QCD calculations~\cite{bib:2}.
While the primordial $J/\psi$ is expected to be dissolved in the deconfined medium, the $J/\psi$ yield is also expected to be enhanced at RHIC energy due
to the abundant creation of $c\bar c$ pairs and the subsequent
recombination of uncorrelated $c\bar c$ pairs in the medium and/or at
the hadronization stage~\cite{bib:3}.

Cold nuclear matter effects (CNM) such as nuclear absorption and gluon shadowing are expected to modify the \JPSI~yield. 
\JPSI~measurement in $d$+Au collisions by PHENIX has shown 
that CNM effects are smaller at RHIC than those observed at SPS energies~\cite{bib:4}.
\vspace{-2mm}
\section{PHENIX Experiment and Data Analysis}
The PHENIX experiment consists of two central arm 
spectrometers, each of which covers the pseudo-rapidity 
range $|\eta| < 0.35$ and 90 degrees in azimuthal angle, and 
two forward spectrometers covering  $1.2<|\eta|<2.4$ with full 
azimuthal acceptance~\cite{bib:phenix}.

The \JPSI~yield is obtained from the unlike-sign dilepton invariant 
mass spectrum after
subtracting combinatorial background using an event mixing method 
for each centrality class, transverse momentum and rapidity bin.
Finally, the numbers of reconstructed $J/\psi$'s are $\sim$1000 
for the di-electron channel 
and $\sim$4500 for the di-muon channel in minimum bias Au+Au collisions.
The invariant $J/\psi$ yield is extracted by correcting the number of
recorded events for the acceptance and efficiency of the spectrometers~\cite{bib:jpsi_auau}. 
The \JPSI~yield measured in 2005 $p+p$ collisions 
at $\sqrt{s}$=200~GeV~\cite{bib:jpsi_pp} was used 
in the calculation of \RAA~for Au+Au collisions.

\section{Results}
\begin{figure}[htbp]
  \begin{center}
  \resizebox{15cm}{!}{\includegraphics{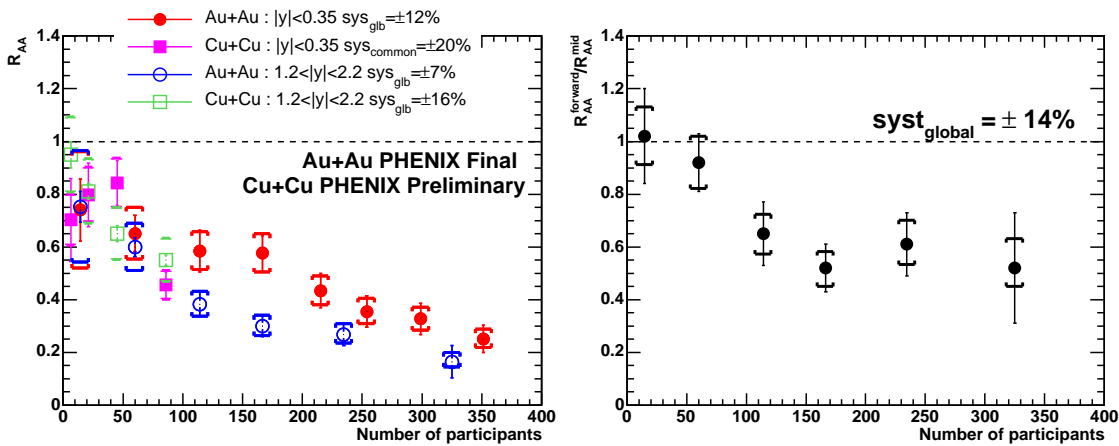}}
  \caption{Left: \RAA~of \JPSI~as a function of the number of participants 
    $N_{part}$ in Au+Au (circle symbols) and Cu+Cu collisions (square symbols) 
    at mid-rapidity (closed symbols) and at forward-rapidity (open symbols).
    Right : Ratio of \RAA~at forward-rapidity to that at mid-rapidity 
    in Au+Au collisions.}
  \label{fig:raa_rhic}
  \end{center}
\end{figure}

Fig.~\ref{fig:raa_rhic} (left) shows \RAA~of \JPSI~as a function of the number of 
participants $N_{part}$ in Au+Au (circle symbols) and Cu+Cu collisions (square symbols) 
at mid-rapidity (closed symbols) and at forward-rapidity (open symbols). 
\RAA~is similar between mid-rapidity and forward-rapidity up to $N_{part}\sim100$ and 
stronger suppression is observed at forward-rapidity for $N_{part}\ge100$. 
Fig.~\ref{fig:raa_rhic} (right) shows the ratio of \RAA~at 
forward-rapidity to that 
at mid-rapidity, which goes down to $\sim$0.6 for $N_{part}\ge100$.
\vspace{-3mm}

\begin{figure}[htbp]
  \centering
  \resizebox{16cm}{!}{\includegraphics{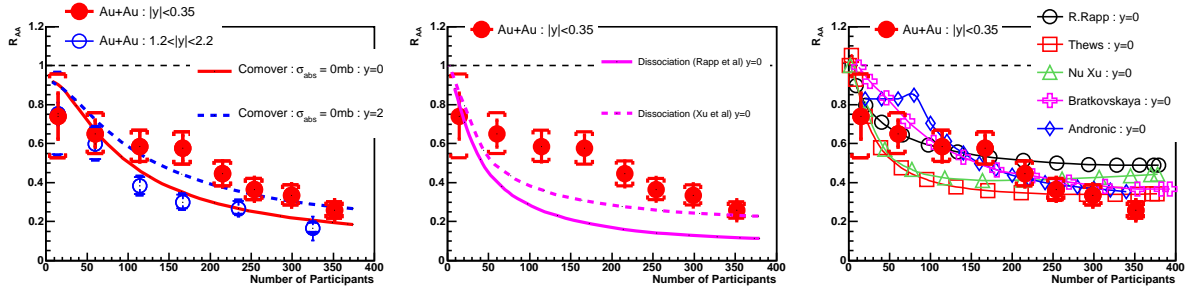}}
  \caption{Left: Comparison of \RAA~to the models with dissociation by comovers. 
    Middle: Comparison of \RAA~to the models with dissociation by thermal gluons. 
    Right: Comparison of \RAA~to the models with dissociation and recombination of \JPSI.}
  \label{fig:raa_comp}
\end{figure}

The left and middle panels of Fig.~\ref{fig:raa_comp} show comparison of 
\RAA~in Au+Au collisions to the models involving only the dissociation 
of \JPSI~by comoving partons and hadrons and by thermal gluons, 
respectively~\cite{bib:capela, bib:rapp1, bib:nuxu1}. 
These models overestimate \JPSI~suppression observed at mid-rapidity at RHIC. 
The predictions, which take into account the recombination of \JPSI~from $c\bar{c}$ pairs in the medium or 
at hadronization stage, are shown in the right panel of 
Fig.~\ref{fig:raa_comp}~\cite{bib:rapp1, bib:nuxu1, bib:thews, bib:anton, bib:brat}.
They match the data better than the models with dissociation only.
However, charm production and its modifications in Au+Au collisions, which are 
input information for recombination scenario, are unclear and need to be understood.
From the experimental side, measurement of \JPSI~azimuthal anisotropy will provide 
useful and direct information on recombination of \JPSI, which will be done in upcoming Au+Au data taking. 
\vspace{-3mm}

\begin{figure}[htbp]
  \centering
  \resizebox{15cm}{!}{\includegraphics{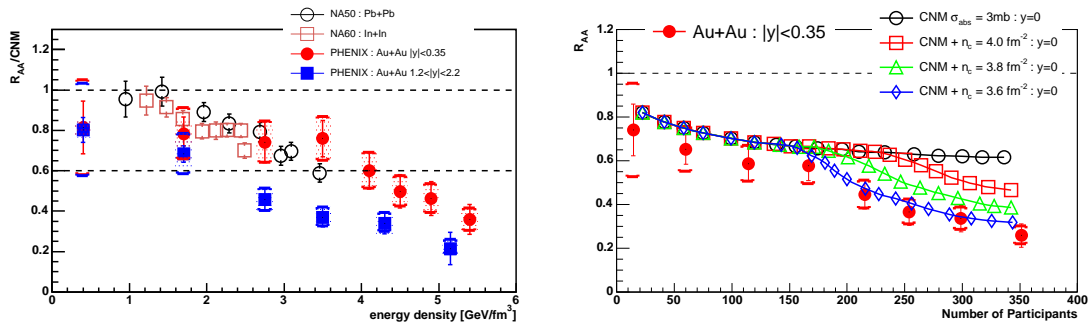}}
  \caption{Left : \RAA/CNM as a function of Bjorken energy density
    in NA50 Pb+Pb collisions (\SNN=17.3~GeV),
    NA60 In+In collisions (\SNN=17.3~GeV) and Au+Au collisions (\SNN=200~GeV),
    where the formation time is assumed to be 1~fm/$c$ for both SPS ans RHIC.}
  \label{fig:saa}
\end{figure}

To extract the final state effects, \RAA~was divided by that expected 
from CNM effects (\RAA/CNM).
CNM effects in Au+Au collisions were extrapolated from those in $d$+Au 
collisions~\cite{bib:vogt}. 
Fig.~\ref{fig:saa} (left) shows \RAA/CNM as a function of Bjorken energy 
density in NA50 Pb+Pb collisions (\SNN=17.3~GeV), 
NA60 In+In collisions (\SNN=17.3~GeV) and Au+Au collisions (\SNN=200~GeV). 
The formation time here is assumed to be 1~fm/$c$ for both SPS and RHIC, which 
could be larger than 1~fm/$c$ at the lower SPS energy and smaller at 
the higher RHIC energy. 
A nuclear absorption cross section of 1 mb was used 
in the calculation of \RAA/CNM for RHIC and the additional systematical uncertainties from CNM effects, which are shown as boxes, were estimated using nuclear
absorption cross sections of 0 mb and 2 mb.
\vspace{-3mm}
\JPSI~suppression at SPS can be interpreted as the melting of only $\chi_{c}$ and $\psi^{\prime}$ since 
they are expected to be dissolved at lower temperature than $J/\psi$ and 
they contribute $\sim$40\% of its total yield 
via decay (feed-down)~\cite{bib:melt}.
It is seen that \JPSI~suppression at RHIC is stronger than the expectation 
from only $\chi_{c}$ and $\psi^{\prime}$ melting in central collisions.
However, the error is too large to 
conclude that direct produced $J/\psi$'s are suppressed at RHIC and 
a more precise measurement of CNM effects is urgently needed. 
Also the fraction of $J/\psi$ from $\chi_{c}$ and $\psi^{\prime}$ 
decay needs to be measured at RHIC energy.
Fig.~\ref{fig:saa} (right) shows the comparison of \RAA to the 
threshold model, which is associated with the onset of 
suppression of directly produced $J/\psi$~\cite{bib:threshold} and 
reproduce the tendency of \JPSI~suppression at mid-rapidity. 

\section{Summary}
PHENIX measured the \JPSI~yield in Au+Au and Cu+Cu collisions at \SNN~=~200~GeV at 
mid-rapidity and forward-rapidity. The stronger suppression is 
observed at forward-rapidity for $N_{part}\ge100$. 
The destruction of \JPSI~by thermal gluons does not
reproduce the observed suppression and dissociation/recombination scenario 
is favored at RHIC energy. 
However, charm production and its modifications in medium are
unclear and need to be understood.
\RAA/CNM at RHIC shows that the \JPSI~suppression seems to be stronger 
than expected 
from the melting of only $\chi_{c}$ and $\psi^{\prime}$ in central collisions. 
However, the error is too large to draw a firm conclusion. 
What should be done in the future experiments is to 
measure CNM effects precisely, the feed-down contribution 
from $\chi_{c}$ and $\psi^{\prime}$ at RHIC energy and 
also the azimuthal anisotropy of \JPSI, which provide more detailed 
information 
to understand the medium effects for \JPSI~production in heavy ion collisions.

\end{document}